\documentstyle[prl,aps]{revtex}
\input{epsf}
\draft
\begin{document}
\twocolumn[\hsize\textwidth\columnwidth\hsize\csname@twocolumnfalse\endcsname
\title{Tunable Pinning of Burst-Waves in Extended Systems with Discrete Sources}

\author{Igor Mitkov, Konstantin Kladko,
and John E. Pearson}
\address{
Center for Nonlinear Studies,
Computational Science Methods Group,
and Theoretical Division\\
Los Alamos  National Laboratory,
Los Alamos, NM 87545
}
\date{\today}
\maketitle

\begin{abstract}
We study the dynamics of waves in a system of diffusively coupled
discrete nonlinear sources. We show that the system exhibits burst
waves which are periodic in a traveling-wave reference frame. We
demonstrate that the burst waves are pinned if the diffusive coupling
is below a critical value. When the coupling crosses the critical
value the system undergoes a depinning instability via a saddle-node
bifurcation, and the wave begins to move.  We obtain the universal
scaling for the mean wave velocity just above threshold.

\end{abstract}

\narrowtext
\vskip1pc]
The effect of discrete source distribution for spatially extended
systems is a problem of interest in such disparate fields as the biophysics
of the calcium release waves in living
cells~\cite{Parker98,Dupont},
pinning in the dislocation motion in crystals~\cite{flachkladko}, breathers
in nonlinear crystal lattices~\cite{flachwillis},
Josephson junction arrays~\cite{floriamazo}, and
charge density waves in one-dimensional strongly correlated electron
systems~\cite{floriamazo,fukuyamalee,coppersmith,bishop}.  In order to
elucidate the effects of discreteness, complex models of calcium
release have been solved numerically~\cite{bugrim97,keizer98,spiro98}.
The simple ``fire-diffuse-fire'' model constructed in~\cite{dawson98}
displays burst (or {\it saltatory}) wave fronts. These fronts either
propagate or they do not exist: they cannot undergo pinning. If a
system with wave pinning consists of diffusively coupled discrete
sources~\cite{fukuyamalee}, the standard approach to its analysis has
been to completely discretize the dynamics.  This is done by replacing
the diffusion term with a difference scheme for the field at the
source sites~\cite{floriamazo,coppersmith}. This simplification
neglects the field structure between sites.  The latter is crucial for
an understanding of the system dynamics and universal behavior near the
pinning/depinning transition.

In this Letter we consider a discrete array
of nonlinear reaction sites embedded in a continuum
in which the reactant diffuses.
We study both analytically and numerically the propagation of
burst waves and their pinning, as well as the universal
behavior of the system near the pinning threshold.
Our model is represented by the following diffusion equation
with discrete nonlinear sources (in $N$-dimensional space):
\begin{equation}
u_t \;=\; D\nabla^2 u \;+\;
\alpha\, d^N \sum_i \delta({\mathbf x}-{\mathbf x}_i)f(u)\;,
\label{eq1}
\end{equation}
where $u$ is a dimensionless concentration, $D$ is the diffusion coefficient,
$\alpha$ is the production rate of the reactant, and $d$ is the
distance between neighboring sites (channels). These sites
are located at ${\mathbf x}_i$'s.
The reaction dynamics are specified by a nonlinear function
$f(u)$. To describe the waves of one stable phase propagating into
another stable phase, we choose the {\it bistable}
reaction dynamics~\cite{ch}.
The simplest example of bistability is given by $f(u) = -u(u-u_0)(u-1)\,$,
with two stable fixed points $u = 0,\,1\,$, and one unstable
fixed point $u = u_0\,$.

In the present Letter we study one-dimensional (1D) wave propagation.
Then, after rescaling $x = \tilde{x}d,\, t = \tilde{t}\alpha^{-1}\,$,
we obtain from Eq.~(\ref{eq1})
(after dropping tildes)
\begin{equation}
u_t \;=\; \beta u_{xx} \;+\; \sum_i \delta(x-i)f(u)\;,
\label{eq2}
\end{equation}
with the effective dimensionless diffusion coefficient
$\beta = D/\alpha d^2\,$.
The system dynamics are
determined by the balance between the dissipation and the local
nonlinear forcing. The latter can be expressed through its
potential as $f = - dF/du\,$. The potential $F(u)$ for a bistable
system possesses two minima (stable fixed points of the reaction),
separated by a maximum (unstable fixed point of the reaction).
For cubic forcing the two minima, $u = 0,\,1\,$, have
equal energy if the potential is symmetric, {\it i.e.} $u_0 = 1/2\,$.
In this case no long-time wave propagation is possible.
The energy source for the motion is the difference between
the depths of the potential minima, as determined by
$\gamma = 1/2 - u_0$ (for small $\gamma$ it is $\sim \gamma$). For
$\gamma>0\,,\, u = 0$ is a local minimum and $u = 1$ a global minimum.
For $\gamma < 0\,$, the dynamics are analogous but the two minima
exchange their roles.
We are left, therefore, with two dimensionless parameters which control
the behavior of our system: $\beta$ and $\gamma\,$.

In Eq.~(\ref{eq2}), the effective diffusion
coefficient $\beta$ is the only measure of how close the system is
to its {\it continuum limit}, $u_t = \beta u_{xx} + f(u)\,$,
which is realized when $\beta \to \infty\,$.
The continuous system possesses traveling-wave kink solutions
propagating from the locally stable minimum to the globally stable minimum
(see, {\it e.g.},~\cite{mikhailov}). 
For small $\gamma$ this solution has the form
$u = 1/2\,[\,1-\tanh(z/2\sqrt{2\beta}\,)\,] + O(\gamma)\,$ (in the traveling
wave coordinate $z = x - Ct$), and the propagation velocity
is $C \sim \gamma\sqrt{\beta}$~\cite{mikhailov,kessler}.

Here we investigate the dynamics for all $\beta$,
including the {\it discrete limit}, when $\beta \ll 1\,$
for which the dynamics are more complicated.
We have performed numerical simulations of Eq.~(\ref{eq2}).
To get rid of the singular form of the dynamics and to ensure
convergence with mesh refinement, we have rewritten
the equation in terms of a nonlocal function
$U(x,t) = \int_0^L \, G(x,x^{\prime})\, u(x^{\prime},t)\,dx\,$,
where $L$ is the
system length and $G$ is a Green-function of 1D Laplace operator, with
no-flux boundary condition $u_x = 0$ at $x = 0$ and Dirichlet
boundary condition $u = 0$ at $x = L\,$. The Green-function has the form
$G(x,x^{\prime}) = (x-x^{\prime})\,H(x-x^{\prime}) + x^{\prime} - L\,$ with
$H$ being a step-function. Equation for $U(x,t)$ was integrated implicitly,
and the inverse Green-function used to restore
$u(x,t)$ needed to calculate the nonlinearity.

\begin{figure}[h]
\rightline{ \epsfxsize = 8.0cm \epsffile{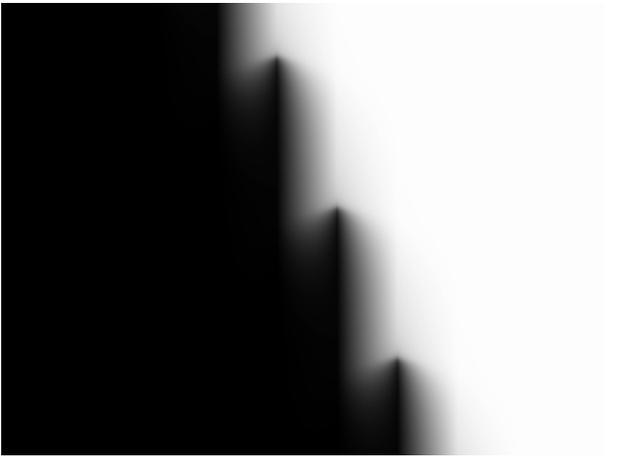}}
\vspace{0.5cm}
\caption{
Space-time plot of $u(x,t)\,$, with black corresponding to $u=1$
and white to $u=0\,$. Time goes positive downward.
Parameters of our model are $\beta = 0.017\,,
\,\gamma = 0.3$ ($u_0 = 0.2$). The system length is $L = 20\,$,
and number of the grid-points is $1000\,$. Time step $d t \approx 0.007\,$.
The period of the bursting is $T = 35.28\,$, total simulation
time is $3 T\,$.
\label{fig1}}
\end{figure}

Far from the continuum limit the system exhibits well developed {\it
burst waves}. Figure~\ref{fig1} shows a space-time plot of
the solution $u(x,t)$ in the burst-wave regime. These burst waves
are propagating solutions which, in the appropriate co-moving reference
frame, are strictly periodic. The bursting period is uniquely
determined by $\beta$ and $\gamma\,$.

For small enough $\beta$ the burst waves are {\it pinned}.
A pinned solution of Eq.~(\ref{eq2})
is a static piece-wise linear curve connecting the stable states $u = 1$ and
$u = 0\,$. The solution of the 1D Laplace equation
is linear, so the sites are the points where the second
derivative is proportional to the $\delta$-function. Thus
the problem is reduced to a set of algebraic equations on the values
of $u$ at the sites
\begin{equation}
\beta\,(u_{i+1} + u_{i-1} - 2 u_i) + f(u_i) \;=\; 0\;,
\label{eq3}
\end{equation}
where $u_i = u(x_i)\,$.

\begin{figure}[h]
\rightline{ \epsfxsize = 8.0cm \epsffile{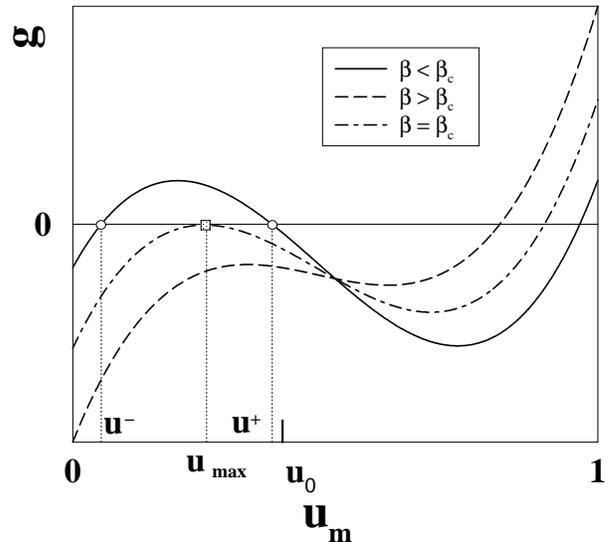}}
\vspace{0.8cm}
\caption{
A graphical representation of $g(u_m)$ used to solve
the stationary problem (\protect\ref{eq3})
to the first order in $\beta\,$.
Parameter $\gamma = 0.1$ ($u_0 = 0.4$).
Stable (unstable) front solution corresponds to
$u_m = u^-\,(u^+)\,$.
\label{fig2}}
\end{figure}

We define the first site (going from $u=1$ to $u=0$)
where the second derivative of $u$ is positive as
``the front'', with corresponding site number
$m\,$. In the discrete limit ($\beta \ll 1$) it can be shown
that the values of $u$ at the sites approach $1$ ($0$) exponentially
with distance from the front. In particular,
$u_{m+i} \sim \beta^i$ and $1 - u_{m-i} \sim \beta^i\,$.
Let us consider the problem of pinning to first order in $\beta\,$.
Then Eq.~(\ref{eq3}) at the front becomes
$g \equiv u_m[\,u_m^2 - u_m (u_0+1) + u_0 + 2\beta\,] - \beta
\,=\, O(\beta^2)\,$. In Figure~\ref{fig2} we plot $g(u_m)$.
By definition of $m\,$,
$(u_m)_{xx} > 0\,$, thus it follows from Eq.~(\ref{eq3}) that
$u_m < u_0\,$. Therefore, the rightmost
root is not appropriate and the existence of solution
depends on the value of $\beta\,$. The critical value of
$\beta\,,\,\beta_c\,$, at which stationary solutions cease to exist,
corresponds to the situation when two remaining roots of
$g(u_m)\,,\,u_m = u^-$ and
$u_m = u^+$ merge at the graph maximum $u = u_{max}\,$.

The stability of a stationary solution of Eq.~(\ref{eq2}), $\;u_i^*\,$
($i = 1,2,\ldots$), is determined by the eigenvalues of the system
obtained by linearizing system~(\ref{eq3}).  This yields the maximum
eigenvalue $\lambda_{max} = f_u(u_m) - 2\beta + O(\beta^2)\,$ which,
for the roots $u_m = u^-$ and $u_m = u^+\,$, is negative and positive,
respectively. Since all remaining eigenvalues are negative, $u = u^+$
is a {\it saddle} and $u = u^-$ is a {\it node} (see Fig.~\ref{fig2}).
If we substitute $\beta = \beta_c$ in the expression for
$\lambda_{max}$ we find that $\lambda_{max} = 0$ [see the definition
of $g(u_m)$].  That is, the stability boundary for the stationary solution
coincides with that of its existence.  This implies that the
pinning/depinning instability occurs via {\it saddle-node
bifurcation}.

\begin{figure}[h]
\rightline{ \epsfxsize = 8.0cm \epsffile{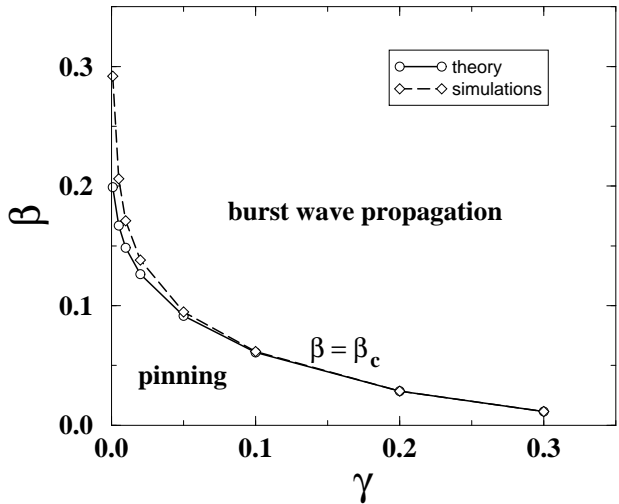}}
\caption{
Bifurcation line $\beta = \beta_c(\gamma)$
for pinning/depinning instability in the plane $\gamma-\beta$ of our model
parameters. In the region above the line burst
waves propagate and below the line they are pinned.
\label{fig3}}
\end{figure}

An analysis similar to the above, but to second order in 
$\beta$, yields the bifurcation line for the pinning/depinning transition
in $(\gamma\,,\,\beta)$ space.
In Figure~\ref{fig3} we compare the analytical bifurcation line with
that obtained by direct numerical simulations of Eq.~(\ref{eq2}).
As shown in the figure, the theory is
in good quantitative agreement with the simulations, even {\it beyond}
the applicability of the small $\beta$ perturbation theory developed above.
For $\gamma \to 0$, the energy source
of wave motion vanishes and consequently $\beta_c$ diverges.

Let us now study the universal behavior of our system near
the critical diffusion coefficient $\beta = \beta_{c}$.
First we note that the system dynamics are {\it variational}, {\it i.e.},
\begin{equation}
\label{E1}
\frac{\partial u}{\partial t} \;=\; - \frac{\delta E[u]}{\delta u},
\end{equation}
where $\delta/\delta u$ is a variational derivative
in the space of functions $u(x)$ of a nonlinear functional
$E[u]$, given by
\begin{equation}
\label{E2}
E[u] \;=\; \frac{\beta}{2} \int \left( \frac{\partial u}{\partial x}\right)^2
dx \;+\; \sum_i F\left[ u(i) \right]\;.
\end{equation}
This functional diverges for infinite systems.
Therefore we use only its difference for different functions $u(x)$.

Denote a small deviation of $\beta$ from the critical value
as $\epsilon = \beta - \beta_{c}$ and consider
first the case of $\epsilon < 0\,$. Then, diffusion
is not strong enough to cause wave propagation, and
there exists a linearly stable stationary kink solution $u_s(x)\,$,
corresponding to $u^-$ in Figure~\ref{fig2}.
Let $u^n(x)$ be a function obtained by translating $u(x)\,$, such that
$u^n(x) = u(x+n)\,$.
If $u_s(x)$ is a minimum of $E[u]$ then $u_s^n(x)$
is also a minimum of $E[u]$ for any integer $n\,$, implying that
the functional $E[u]$ has an infinite set of minima.
One can show that $E[u^1_s(x)] - E[u_s(x)] = F(u\equiv 1) - F(u\equiv 0)\,$.

Consider two adjacent minima of $E[u]\,,\,u_s(x)$ and $u^1_s(x)\,$.
The corresponding basins of attractions have a common boundary.
A minimum of $E[u]$ on the boundary is
a saddle point $u_u(x)$ of the functional $E[u]\,$,
which represents an unstable kink,
corresponding to $u^+$ in Figure~\ref{fig2}.
Due to the discrete translational symmetry of
the system, $u^n_u(x)$ is also a saddle point of $E[u]$ for any integer $n$.
A variation of the functional $E[u]$ in a vicinity of the unstable kink
$u_u(x)$ is represented by a quadratic form (in the variation of the
function), having one negative eigenvalue
and a corresponding unstable direction $\bf{n}\,$. There exist two separatrix
lines of $E[u]\,,\, L_1$ and $L_2\,$, which start at $u_u$ and go
to $u_s$ and $u^1_s\,$, respectively.
The tangent directions of the separatrices
at $u_u$ are collinear to $\bf{n}\,$. The curve $L_{12}= L_1 \bigcup L_2$ joins
two stable minima $u_s$ and $u^1_s$ and contains the saddle $u_u\,$. If
we continue this curve infinitely it will contain all nodes
and saddles of $E[u]\,$.

\begin{figure}[h]
\rightline{ \epsfxsize = 7.0cm \epsffile{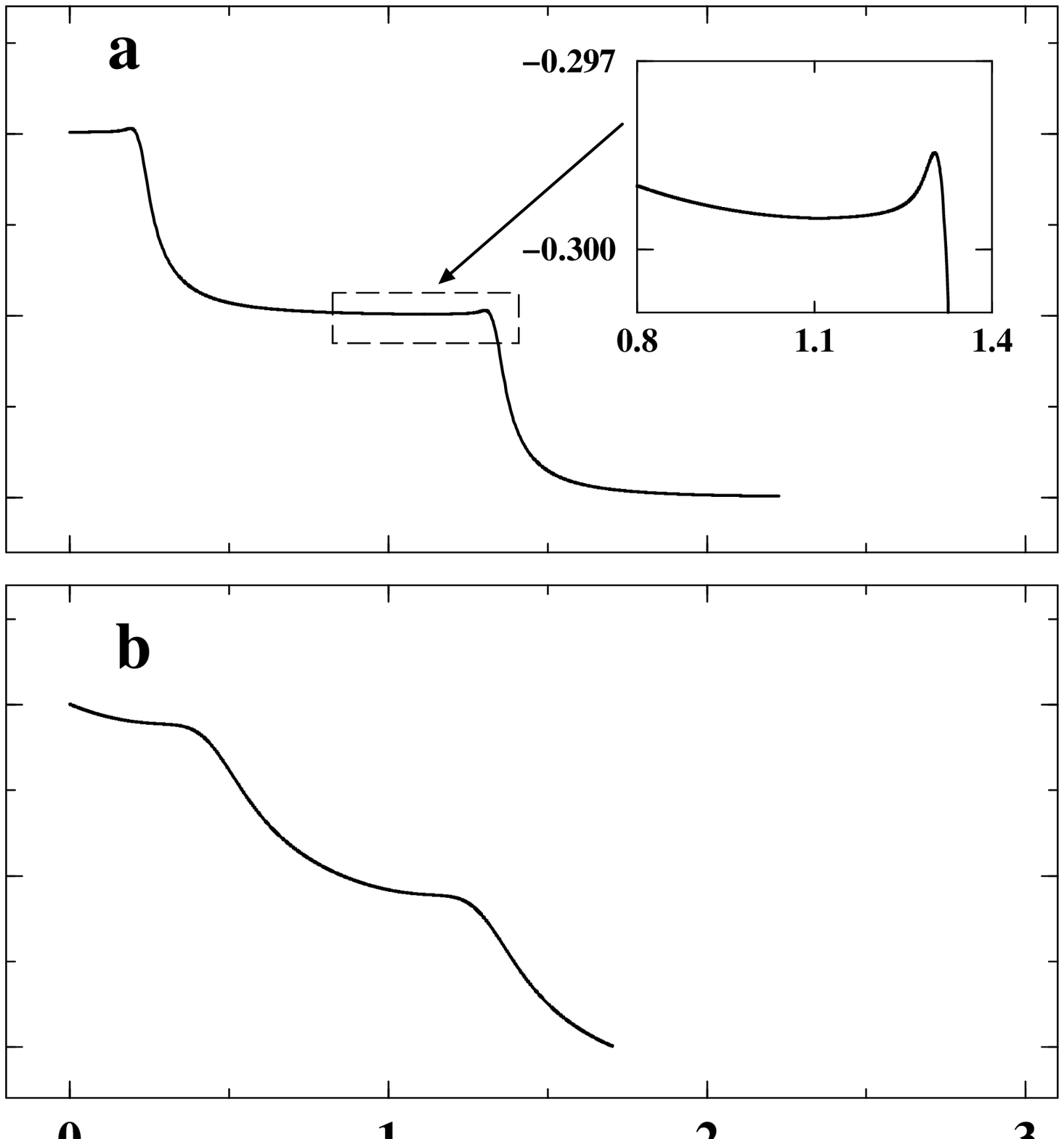}}
\vspace{0.8cm}
\caption{
Dependence of the energy functional $E$ upon the arclength $s$
along the trajectory in $u$-space. (a) $\epsilon < 0\,$;
parameters of the simulations are $\beta = 0.001\,,
\,\gamma = 0.3\,,\, L = 10\,$,
number of grid-points is $1000\,$, time step $d t = 0.01\,$.
Inset shows the structure of saddle-node pair.
(b) $\epsilon > 0\,$; parameters are the same as in Fig.~\protect\ref{fig1}.
\label{fig4}}
\end{figure}

Let us introduce a normal coordinate (arclength) $s$ on this curve,
$ds^2 = dt^2 \int (\partial u/\partial t)^2 dx\,$.
Figure~\ref{fig4}(a)
shows $E(s)$ obtained numerically for $\epsilon < 0\,$.
The minima and maxima of $E(s)$ [see inset in Fig.~\ref{fig4}(a)]
correspond to stable kinks $u_s$ and unstable kinks $u_u\,$, respectively.
It can be shown, using the definitions of $E$ and $s$, that
the system dynamics on the trajectory is governed by

\begin{equation}
\label{E3}
\frac{ds}{dt} \;=\; - \frac{dE}{ds}\;.
\end{equation}

As $\epsilon$ increases through $0$,
adjacent maxima and minima of $E(s)$ merge,
a saddle-node bifurcation occurs,
and the kink begins to move. The function $E(s)$
for this case is plotted in Fig.~\ref{fig4}(b).
At $\epsilon = 0\,$, the extrema coalesce in inflection points,
where the first and the second derivatives are zero.
The behavior of $E(s)$ near this
point is given by $E(s) = E(0) -  As^3/3$, where $A>0\,$. For
$\epsilon > 0$ the expansion of $E(s)$ acquires a linear in $s$ part
($\sim\epsilon$), so we can write
$E(s) = E(0) - A s^3/3 - B \epsilon s\,$.
We can estimate the period of the kink motion just above
the bifurcation as
\begin{equation}
\label{E4}
T \sim  \int_0^{s_0}  \left|\frac{dt}{ds}\right| ds =
\int_0^{s_0} \frac{1}{B \epsilon + A s^2}\; ds\;,
\end{equation}
where $s_0$ is a finite constant.

\begin{figure}[h]
\rightline{ \epsfxsize = 8.0cm \epsffile{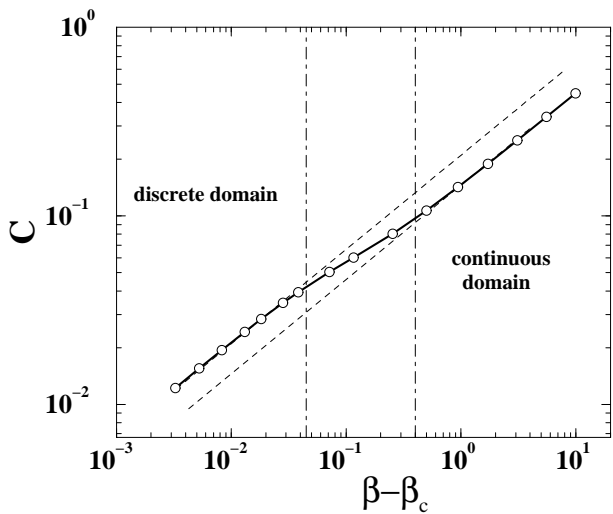}}
\caption{
Wave propagation velocity $C$ vs $\beta-\beta_c$
(solid line). Dashed lines with slope $1/2$ show the scalings
$C \sim \sqrt{\epsilon}=\sqrt{\beta-\beta_c}$ and $C \sim \sqrt{\beta}$
for discrete and continuous domain, respectively.
Dashed-dotted lines illustrate
the boundaries of continuous, discrete, and transient
domains of $\beta\,$. The parameters are $\gamma=0.1$ and
$\beta_c\approx 0.0617$.
\label{fig5}}
\end{figure}

For small $\epsilon\,$, $T$ in (\ref{E4}) does not depend on $s_0$
and is
$\sim 1/\sqrt{\epsilon}\,$. Therefore, the average kink velocity is
$C = 1/T\sim \sqrt{\epsilon}$ near the bifurcation point.
Fig.~\ref{fig5} is a plot of the velocity $C$ as a function of
$\epsilon = \beta-\beta_c$ obtained by numerical simulations of
Eq.~(\ref{eq2}).
The $\sqrt{\epsilon}$-scaling near the critical point is clearly seen.
For finite $\epsilon\,$, the scaling breaks and, after a transient
range of $\beta\,$, the expected continuum scaling $C \sim\sqrt{\beta}$
appears.

If we identify the set of points $x+n$
as one point, a moving kink will be a periodic trajectory (limit cycle)
on the reduced manifold, born by a saddle-node bifurcation.
This type of behavior is well known for
finite-dimensional systems ~\cite{arnold}. A simple exactly soluble
example which has
both the symmetries and the scaling behavior discussed here is given by
$dx/dt = 1 + (1-\epsilon) \cos{2 \pi x}$. For $\epsilon < 0$ the system has
an infinity of fixed points and for $\epsilon > 0$ the fixed points cease to
exist and the solution propagates with $x(t+T)=x(t)+1$
and $T \sim 1/\sqrt{\epsilon}$ for small $\epsilon$.

We have shown that there exist standing kink solutions for
one-component bistable reaction-diffusion systems with discrete
sources. We have also obtained the universal scaling for the mean
velocity just above the pinning threshold. We conjecture that pinning
is universal in all bistable systems with discrete sources.  Pinning
was also observed in numerical simulations of a nonvariational but
bistable model for the calcium release wave in cardiac
myocytes~\cite{spiro98}.  These results are in contrast to those
in \cite{keizer98} and \cite{dawson98} in which saltatory or burst
waves were found with the velocity scaling as $C \sim D/d$ for small
$\beta\,$, but no pinning.
The systems in \cite{keizer98} and \cite{dawson98} were not
bistable which is consistent with our conjecture.
In all models studied so far, waves fail to propagate for
large enough site spacing. The manner in which propagation failure
occurs differs from system to system. For example in the
``fire-diffuse-fire'' model~\cite{dawson98} of calcium dynamics,
propagation failure occurs through a sequence of period-doubling
bifurcations and crises.  These facts raise an interesting open
question. In what ways can waves fail to propagate as the source
spacing is increased? To this end it would be useful to systematically
generalize the
work here to nonvariational systems. Such systems would be also more
realistic as models of calcium dynamics in living cells.

We are grateful to I. Aranson, S. Ponce Dawson, B. Ermentrout,
E. Ben-Naim, and K. Vixie for fruitful discussions.
This work was supported by the LDRD grant XAA7 at Los Alamos National
Laboratory.

\end{document}